\documentclass[11pt]{article}
\usepackage{amsmath,amssymb,amsfonts,bbm}

\newcommand{\be}{\begin{equation}}
\newcommand{\ee}{\end{equation}}
\newcommand{\bea}{\begin{eqnarray}}
\newcommand{\eea}{\end{eqnarray}}
\newcommand{\ba}{\begin{array}}
\newcommand{\ea}{\end{array}}

\newcommand{\N}{\mathcal{N}}

\long\def\symbolfootnote[#1]#2{\begingroup%
\def\thefootnote{\fnsymbol{footnote}}\footnote[#1]{#2}\endgroup}
\begin{document}

\thispagestyle{empty}\vspace{40pt}

\hfill{}

\vspace{128pt}

\begin{center}
    \textbf{\Large Split-complex representation \\ of the universal hypermultiplet}\\
    \vspace{40pt}

    Moataz H. Emam\symbolfootnote[1]{\tt moataz.emam@cortland.edu}

    \vspace{12pt}   \textit{Department of Physics}\\
                    \textit{SUNY College at Cortland}\\
                    \textit{Cortland, NY 13045, USA}\\
\end{center}

\vspace{40pt}

\begin{abstract}

Split-complex fields usually appear in the context of Euclidean supersymmetry. In this paper, we propose that this can be generalized to the non-Euclidean case and that, in fact, the split-complex representation may be the most natural way to formulate the scalar fields of the five dimensional universal hypermultiplet. We supplement earlier evidence of this by studying a specific class of solutions and explicitly showing that it seems to favor this formulation. We also argue that this is directly related to the symplectic structure of the general hypermultiplet fields arising from non-trivial Calabi-Yau moduli. As part of the argument, we find new explicit instanton and 3-brane solutions coupled to the four scalar fields of the universal hypermultiplet.

\end{abstract}

\newpage



\vspace{15pt}

\pagebreak

\section{Introduction}

Studying the dimensional reduction of M-theory and the theories that arise from it is an ongoing quest of great relevance to further understanding the general picture of the string theory landscape. Particularly important are the $\N=2$ (and higher) theories that appear as a result of dimensionally reducing M/string theory over manifolds with special holonomy. These theories contain tensor and vector supersymmetric multiplets, as well as scalar super-fields known as the hypermultiplets. The physics of those fields depend greatly on the inherent symmetries and topology of the underlying sub-space, hence a study of them constitutes one way of studying the sub-manifolds themselves. Such theories in both $D=4$ and $D=5$ are related via certain mathematical transformations such as the so-called c-map, making a study of one essentially the same as the study of another. Our focus in this paper, one in a series, is the study of the five dimensional $\N=2$ supergravity theory that arises from the dimensional reduction of the unique eleven dimensional theory over a Calabi-Yau 3-fold (CY). This contains both vector multiplets and scalar hypermultiplets. An abundance of work on the theory (and the related one in four dimensions) exist in the literature. One notes, however, that most of these study couplings with the vector multiplets sector (for example, see \cite{Behrndt:1997fq,Sabra:1997kq,Sabra:1997dh,Behrndt:1997ny,Sabra:1997yd,Behrndt:1997as,Behrndt:1998eq,Behrndt:1998ns}). In contrast, hypermultiplets-coupled solutions are quite rare (\textit{e.g.} \cite{Behrndt:2000km,Ceresole:2001wi,Behrndt:2002ee,Cacciatori:2002qx}). Hence, our focus on the latter also fills a particular gap in the literature.

The hypermultiplet fields have a rich symplectic structure that arises from the topology of the underlying CY space. We have focused on this in previous works (\cite{Emam:2004nc}, \cite{Emam:2006sr}, \cite{Emam:2009xj}), and have proposed that a certain formulation of the theory, based on defining a symplectic vector space where the hypermultiplet fields appear as symplectic vectors and scalars, is an optimal formulation providing tools for generating solutions and a further understanding of the theory's structure. If one, however, studies only the special case of the universal hypermultiplet (UH), which arises even in the most trivial case of a rigid CY sub-manifold (\cite{Emam:2005bh}, \cite{Emam:2007qa}), the symplectic structure seems to be hidden. In this paper, we propose that this is not quite true, and that the symplectic structure of the more general theory continues to manifest itself if one chooses a formulation of the theory that is based on the so-called ``split-complex'' numbers, as opposed to the usual, traditional, formulation based on the ordinary complex numbers. There is early evidence of this, such as in \cite{Cortes:2005uq} and our \cite{Emam:2007qa}. We further supplement this evidence by studying a class of solutions that satisfy the Bogomol'nyi-Prasad-Sommerfield (BPS) condition if and only if they are written in split-complex form. Specifically, these have the interpretation of instantons and 3-branes in $D=5$.

\section{Complex and split-complex $\N=2$ supergravity}

The dimensional reduction (see \cite{Emam:2010kt} for a review) of $D=11$ supergravity theory over a rigid Calabi-Yau 3-fold with constant K\"{a}hler and complex structure moduli yields an ungauged $\N=2$ supersymmetric gravity theory in $D=5$ with a matter sector comprised of four scalar fields and their superpartners; collectively known as the universal hypermultiplet. These are the dilaton $\sigma$ (volume modulus of the CY space), the universal axion $a$, the pseudo-scalar axion $\chi$ and its complex conjugate $\bar \chi$, all together parameterizing the quaternionic manifold $SU(2,1)/U(2)$ \cite{Ferrara:1989ik, Cecotti:1988qn}. The bosonic part of the action can be written in the following way
\be
    S_5  = \int\limits_5 {\left[ {R\star \mathbf{1} - \frac{1}{2}d\sigma  \wedge \star d\sigma  - e^\sigma  d\chi  \wedge \star d\bar \chi  - \frac{1}{2} e^{2\sigma } \left( {da + \frac{i}{2}f} \right) \wedge \star \left( {da - \frac{i}{2}\bar f} \right)} \right]},\label{ActionCOMPLEX}
\ee
where $\star$ is the $D=5$ Hodge duality operator and we have defined
\bea
    f &=& \left( {\chi d\bar \chi  - \bar \chi d\chi } \right) \nonumber\\
    \bar f &=&  - f,\label{f_definition}
\eea
for brevity. The variation of the action yields the following field equations for $\sigma$, $\left(\chi, \bar\chi\right)$ and $a$ respectively
\bea
    \left( {\Delta \sigma } \right)\star \mathbf{1} - e^\sigma  d\chi  \wedge \star d\bar \chi  - e^{2\sigma } \left( {da + \frac{i}{2}f} \right) \wedge \star\left( { da - \frac{i}{2}\bar f} \right) &=& 0 \label{Dilaton eomCOMPLEX}\\
    d^{\dagger} \left[ {e^\sigma  d\chi  + ie^{2\sigma } \chi \left( {da + \frac{i}{2}f} \right)} \right] &=& 0 \label{Chi eomCOMPLEX}\\
    d^{\dagger} \left[ {e^\sigma  d\bar \chi  - ie^{2\sigma } \bar \chi \left( {da - \frac{i}{2}\bar f} \right)} \right] &=& 0\label{Chi bar eomCOMPLEX}\\
    d^{\dagger} \left[ {e^{2\sigma } \left( {da + \frac{i}{2}f} \right)} \right] &=& 0\label{a eomCOMPLEX},
\eea
where $d^\dag$ is the adjoint exterior derivative and $\Delta$ is the Laplace-De Rahm operator. The full action is invariant under the following set of supersymmetry (SUSY) transformations of the gravitini $\psi$ and hyperini $\xi$ fermionic fields respectively $\left( {M = 0, \cdots ,4} \right)$:
\bea
    \delta _\epsilon  \psi ^1  &=& D\epsilon _1  + \frac{i}{4}e^{  \sigma } \left( {da + \frac{i}{2}f} \right)\epsilon _1  - \frac{1}{4} {{e^{\frac{\sigma }{2}} }}{{ }}d\chi \epsilon _2  \label{1COMPLEX}\\
    \delta _\epsilon  \psi ^2  &=& D\epsilon _2  - \frac{i}{4}e^{  \sigma } \left( {da - \frac{i}{2}\bar f} \right)\epsilon _2  + \frac{1}{4}{{e^{\frac{\sigma }{2}} }}{{ }}d\bar \chi \epsilon _1 \label{Gravitini variationsCOMPLEX}\\
    \delta _\epsilon  \xi _1  &=& \frac{1}{2}\left[ {\left( {\partial _M \sigma } \right) - ie^{  \sigma } \left( {\partial _M a + \frac{i}{2}f_M } \right)} \right]\Gamma ^M \epsilon _1  + \frac{{e^{\frac{\sigma }{2}} }}{{\sqrt 2 }}\left( {\partial _M \chi } \right)\Gamma ^M \epsilon _2  \label{2COMPLEX}\\
    \delta _\epsilon  \xi _2  &=& \frac{1}{2}\left[ {\left( {\partial _M \sigma } \right) + ie^{  \sigma } \left( {\partial _M a - \frac{i}{2}\bar f_M } \right)} \right]\Gamma ^M \epsilon _2  - \frac{{e^{\frac{\sigma }{2}} }}{{\sqrt 2 }}\left( {\partial _M \bar \chi } \right)\Gamma ^M \epsilon _1,
    \label{Hyperini variationsCOMPLEX}
\eea
where
\be
    D = dx^M \left( {\partial _M  + \frac{1}{4}\omega _M^{\,\,\,\,\,\hat M\hat N} \Gamma _{\hat M\hat N} },\label{Covariant derivative}
    \right)
\ee
is the usual covariant derivative, the $\Gamma$'s are the $D=5$ Dirac matrices, $\left(\epsilon_1,\epsilon_2\right)$ are the $\N=2$ SUSY spinors, $\omega$ is the spin connection and the hatted indices are frame indices in a flat tangent space. To tie in to previous work, we recall that the universal axion $a$ is magnetically dual to a 3-form gauge potential $A$, such that its field strength $F=dA$ can be defined as follows
\be
    F = e^{2\sigma } \star\left( {da + \frac{i}{2}f} \right)\label{F_definition}.
\ee
It is interesting to note that $F=0$ does not \emph{necessarily} imply the vanishing of $da$, since the value of $f$ may not be zero.

At this point, we note that there is a strange discrepancy between equations (\ref{1COMPLEX}) and (\ref{Gravitini variationsCOMPLEX}), and then again between (\ref{2COMPLEX}) and (\ref{Hyperini variationsCOMPLEX}) manifest in the appearance of a change of sign in the last term of each pair. As such, the $\left(\psi^1, \psi^2\right)$ and $\left(\xi _1, \xi _2\right)$ variations are not, strictly speaking, complex conjugate of each other. This particular property will result, as we will show, in certain restrictions on the solutions forcing us to resort to a different representation of the axion fields $\left(\chi, \bar\chi\right)$; that of the split-complex numbers rather than the ordinary complex numbers. The split-complex numbers have been previously shown to naturally arise in the context of hypermultiplet couplings obeying Euclidean supersymmetry (see \emph{e.g.} \cite{Cortes:2005uq} and the references within). Furthermore, in a previous paper \cite{Emam:2007qa}, we used the split-complex representation to overcome the problem that the axion fields, for the specific solution we studied therein, had to satisfy
\be
   d\chi  \wedge d\bar \chi  = d\chi  \wedge \star d\bar \chi  = 0,\label{Nilpotent}
\ee
while not themselves explicitly vanishing. One is then led to wonder if the split-complex representation of the axions may not generally be the best, or more `natural', representation to use. To do so, we redefine the axions as follows
\bea
    \chi  &=& \chi _1  + j\chi _2  \nonumber\\
    \bar \chi  &=& \chi _1  - j\chi _2,
\eea
where $\left(\chi _1, \chi _2\right)$ are real functions and the ``imaginary'' number $j$ is defined by $j^2=+1$ but is \emph{not} equal to $\pm1$. Split-complex numbers\footnote{Also known as `para-complex numbers', `real tessarines', `algebraic motors', `hyperbolic complex numbers', `double numbers', `perplex numbers', `Lorentz numbers', and several others.} are a generalization of the ordinary complex numbers satisfying the `hyperbolic scalar product'\footnote{Technically (\ref{Hyperbolic scalar product}) is not a true scalar product because it is not positive definite, but it is this very property that makes it useful as far as our objectives are concerned.}:
\be
    \left| \chi  \right|^2  = \chi _1^2  - \chi _2^2\label{Hyperbolic scalar product}.
\ee

 In contrast to the complex numbers, which form a field, the split-complex numbers form a ring. They have the interesting property, absent from the complex numbers, of containing non-trivial idempotents (other than 0 and 1), where an idempotent $Z$ is defined by $Z^2=Z$. This property can be used to define the so-called diagonal, or null, basis:
\bea
    g &=& \frac{1}{2}\left( {1 + j} \right) \nonumber\\
    \bar g &=& \frac{1}{2}\left( {1 - j} \right),
\eea
such that any split-complex quantity, such as our axion fields, can be written in the form:
\bea
    \chi  &=& \left( {\chi _1  + \chi _2 } \right)g + \left( {\chi _1  - \chi _2 } \right)\bar g \nonumber\\
    \bar \chi  &=& \left( {\chi _1  - \chi _2 } \right)g + \left( {\chi _1  + \chi _2 } \right)\bar g,\label{Split complex representation}
\eea
where $g$ is idempotent as well as null, \emph{i.e.} $\left| g \right|^2  = g\bar g = 0$. It is somewhat straightforward to see that (\ref{Hyperbolic scalar product}), being the defining relation of split-complex mathematics, naturally arises from the symplectic scalar product of the full set of hypermultiplets, detailed in \cite{Emam:2009xj}, because it has the same form. And as such, it may not be too surprising that it is, as we argue, the better representation for the theory. The bosonic action, field equations and SUSY variations can then be rewritten ($i\rightarrow j$):
\be
    S_5  = \int\limits_5 {\left[ {R\star \mathbf{1} - \frac{1}{2}d\sigma  \wedge \star d\sigma  - e^\sigma  d\chi  \wedge \star d\bar \chi  - \frac{1}{2} e^{2\sigma } \left( {da + \frac{j}{2}f} \right) \wedge \star \left( {da - \frac{j}{2}\bar f} \right)} \right]},\label{ActionSPLIT}
\ee
\bea
    \left( {\Delta \sigma } \right)\star \mathbf{1} - e^\sigma  d\chi  \wedge \star d\bar \chi  - e^{2\sigma } \left( {da + \frac{j}{2}f} \right) \wedge \star\left( { da - \frac{j}{2}\bar f} \right) &=& 0 \label{Dilaton eomSPLIT}\\
    d^{\dagger} \left[ {e^\sigma  d\chi  + je^{2\sigma } \chi \left( {da + \frac{j}{2}f} \right)} \right] &=& 0 \label{Chi eomSPLIT}\\
    d^{\dagger} \left[ {e^\sigma  d\bar \chi  - je^{2\sigma } \bar \chi \left( {da - \frac{j}{2}\bar f} \right)} \right] &=& 0\label{Chi bar eomSPLIT}\\
    d^{\dagger} \left[ {e^{2\sigma } \left( {da + \frac{j}{2}f} \right)} \right] &=& 0\label{a eomSPLIT},
\eea
\bea
    \delta _\epsilon  \psi ^1  &=& D\epsilon _1  + \frac{j}{4}e^{  \sigma } \left( {da + \frac{j}{2}f} \right)\epsilon _1  - \frac{j}{4} {{e^{\frac{\sigma }{2}} }}{{ }}d\chi \epsilon _2  \label{1SPLIT}\\
    \delta _\epsilon  \psi ^2  &=& D\epsilon _2  - \frac{j}{4}e^{  \sigma } \left( {da - \frac{j}{2}\bar f} \right)\epsilon _2  + \frac{j}{4}{{e^{\frac{\sigma }{2}} }}{{ }}d\bar \chi \epsilon _1 \label{Gravitini variationsSPLIT}\\
    \delta _\epsilon  \xi _1  &=& \frac{1}{2}\left[ {\left( {\partial _M \sigma } \right) - je^{  \sigma } \left( {\partial _M a + \frac{j}{2}f_M } \right)} \right]\Gamma ^M \epsilon _1  + j\frac{{e^{\frac{\sigma }{2}} }}{{\sqrt 2 }}\left( {\partial _M \chi } \right)\Gamma ^M \epsilon _2  \label{2SPLIT}\\
    \delta _\epsilon  \xi _2  &=& \frac{1}{2}\left[ {\left( {\partial _M \sigma } \right) + je^{  \sigma } \left( {\partial _M a - \frac{j}{2}\bar f_M } \right)} \right]\Gamma ^M \epsilon _2  - j\frac{{e^{\frac{\sigma }{2}} }}{{\sqrt 2 }}\left( {\partial _M \bar \chi } \right)\Gamma ^M \epsilon _1.
    \label{Hyperini variationsSPLIT}
\eea

Since inserting ``$j$'' in by hand is tantamount to simply inserting a factor of ``1'', we have done so in the last term of the SUSY variations to fix the aforementioned discrepancy by making them conjugate of each other; clearly an advantage over the traditional (ordinary) complex formulation.

\section{Spacetime background}

We will study the possible complex/split-complex universal hypermultiplet solutions in the background of a $p$-brane spacetime with (Poincar\'{e})$_{p+1}\times SO\left( 4-p \right)$ symmetry, including instantons ($p=-1$). The most general such metric is
\be
    ds^2  = e^{2C\sigma  \left( r \right) } \eta _{ab} dx^a dx^b  + e^{2B\sigma\left( r \right)} \delta _{\mu \nu } dx^\mu  dx^\nu  ,\,\,\,\,\,\,\,\,\,\,\,\,\,\,\,\,a,b = 0,...,p\,\,\,\,\,\,\,\,\,\,\,\,\,\,\,\,\mu ,\nu  = p + 1,...,4, \label{pbrane metric}
\ee
where $C $ and $B$ are constants, $(x^1,...,x^p)$ define the directions tangent to the brane and $(x^{p+1},...,x^4)$ those transverse
to it. The variable $r = \left( {\delta ^{\mu \nu } x_\mu x_\nu  } \right)^{{1 \mathord{\left/  {\vphantom {1 2}} \right.  \kern-\nulldelimiterspace} 2}} $ is the usual radial variable in the directions orthogonal to the brane. If one requires the brane to satisfy the Bogomol'nyi-Prasad-Sommerfield condition, breaking half of the supersymmetries of the theory, then one must also necessarily require the vanishing of the variation of gravitini and hyperini backgrounds, \textit{i.e.} $\delta \psi=0$ and $\delta \xi=0$, which, we will show, suffers from problems if the theory is in the ordinary complex representation. Since it turns out that both $\delta \psi=0$ and the Einstein equations require the vanishing of the constant $C$, we set it to zero from the start:
\begin{equation}\label{pbrane metric with C=0}
    ds^2  = \eta _{ab} dx^a dx^b  + e^{2B\sigma\left( r \right)   } \delta _{\mu \nu } dx^\mu  dx^\nu.
\end{equation}

We find the following components of the Einstein and stress tensors respectively
\bea
 G_{ab}  &=& B\left( {3 - p} \right)\eta _{ab} g ^{\mu \nu } \left( {\partial _\mu  \partial _\nu  \sigma } \right) + \frac{1}{2}B^2 \left( {2 - p} \right)\left( {3 - p} \right)\eta _{ab}  \left( {\partial _\alpha  \sigma } \right)\left( {\partial ^\alpha  \sigma } \right) \nonumber\\
 G_{\mu \nu }  &=&  - B\left( {2 - p} \right)\left( {\partial _\mu  \partial _\nu  \sigma } \right) + B\left( {2 - p} \right)\delta _{\mu \nu } \delta ^{\alpha \beta } \left( {\partial _\alpha  \partial _\beta  \sigma } \right) \nonumber\\
& &  + \frac{1}{2}B^2 \left( {1 - p} \right)\left( {2 - p} \right)g _{\mu \nu }  \left( {\partial _\alpha  \sigma } \right)\left( {\partial ^\alpha  \sigma } \right) + B^2 \left( {2 - p} \right)\left( {\partial _\mu  \sigma } \right)\left( {\partial _\nu  \sigma } \right).\label{Einstein tensor}
\eea
\bea
 T_{ab}  &=&  \frac{1}{4}\eta _{ab} \left( {\partial _\alpha  \sigma } \right)\left( {\partial ^\alpha  \sigma } \right)  +\frac{1}{2}\eta _{ab} e^\sigma  \left( {\partial _\alpha  \chi } \right)\left( {\partial ^\alpha  \bar \chi } \right)\nonumber\\
 &+&  \frac{1}{4}\eta _{ab} e^{2\sigma } \left( {\partial _\alpha a + \frac{c}{2}f_\alpha } \right)\left( {\partial ^\alpha a - \frac{c}{2}\bar f^\alpha } \right)  \nonumber\\
 T_{\mu \nu }  &=&  \frac{1}{4}g _{\mu \nu } \left( {\partial _\alpha  \sigma } \right)\left( {\partial ^\alpha  \sigma } \right) -\frac{1}{2}\left( {\partial _\mu  \sigma } \right)\left( {\partial _\nu  \sigma } \right)\nonumber\\
  &+&    \frac{1}{2}e^{\sigma } g _{\mu \nu }   \left( {\partial _\alpha  \chi } \right)\left( {\partial ^\alpha  \bar \chi } \right)-e^\sigma  \left( {\partial _\mu  \chi } \right)\left( {\partial _\nu  \bar \chi } \right) \nonumber\\
  &+&   \frac{1}{4}e^{2\sigma } g _{\mu \nu } \left( {\partial _\alpha a + \frac{c}{2}f_\alpha } \right)\left( {\partial ^\alpha a - \frac{c}{2}\bar f^\alpha } \right) - \frac{1}{2}e^{2\sigma } \left( {\partial _\mu a + \frac{c}{2}f_\mu } \right)\left( {\partial _\nu a - \frac{c}{2}\bar f_\nu } \right),\label{Stress tensor}
\eea
where $c$ is either $i$ or $j$, depending on which representation we choose to use. Other bits and pieces needed for the calculations are as follows: The f\"{u}nfbeins and Christoffel symbols are:
\begin{eqnarray}
    e_{\;\;b}^{\hat a}  &=& \eta _b^{\hat a},\quad \quad \quad \quad
    e_{\;\;\nu}    ^{\hat     \mu }  = e^{B\sigma } \delta _\nu ^{\hat \mu }  \nonumber \\
    \Gamma _{\nu \rho }^\mu   &=& B\left[ {\delta _\nu ^\mu  \left( {\partial _\rho  \sigma
    }     \right) + \delta _\rho ^\mu  \left( {\partial _\nu  \sigma } \right) - \delta
    _{\nu \rho } \delta ^{\mu \alpha } \left( {\partial _\alpha  \sigma } \right)}
    \right],
\end{eqnarray}
resulting in the following spin connections and covariant derivatives:
\begin{eqnarray}
    \omega _\alpha ^{\;\;\hat \beta \hat \gamma }&=& B\left( {\delta _\alpha ^{\hat \beta }
    \delta ^{\hat \gamma \rho }  - \delta ^{\hat \beta \rho } \delta _\alpha ^{\hat \gamma
    }     } \right)\left( {\partial _\rho  \sigma } \right)\nonumber \\
     D_a   &=& \partial _a  \nonumber \\
    D_\mu   &=& \partial _\mu   + \frac{B}{2}\left( {\partial _\nu  \sigma
    }     \right){\Gamma _\mu } ^\nu,
\end{eqnarray}
as well as the Dirac matrices projection conditions\footnote{The Einstein summation convention is \emph{not} used over the index
$s$.}:
\begin{eqnarray}
    \Gamma _{\hat \mu \hat \nu } \epsilon _s  &= b_s \varepsilon _{\hat
    \mu     \hat \nu } \epsilon _s,  \quad\quad\quad\quad s &=(1,2),\quad b_s=\pm
    c \nonumber \\
    \Gamma _\mu^{\;\;\;\nu} \epsilon _s  &= b_s {\varepsilon_\mu}^{\;\nu}\epsilon
    _s, \quad\quad
    \Gamma ^\mu  \epsilon _s  &=  - b_s {\varepsilon_\nu}^{\;\mu} \Gamma
    ^\nu          \epsilon _s.\label{projection}
\end{eqnarray}

\section{Analysis}

We begin by considering equation (\ref{a eomCOMPLEX}/\ref{a eomSPLIT}). It can be integrated once to yield
\be
    e^{2\sigma } \left( {da + \frac{c}{2}f} \right) = \alpha dH,\label{BASIC ANSATZ}
\ee
where $H\left(r\right)$ is a harmonic function; $\Delta H = d^\dagger dH = 0$. The reality of the constant $\alpha$ is required for all positive values of $p$, however it may become complex for the case $B=0$, since this gives a Minkowski spacetime background, which, upon performing a Wick, or Wick-like, rotation $t\to cx^0 $ yields Euclidean spacetime. This automatically restores the reality of $da$ and clearly yields an instanton solution to the theory.

Since we require that $\delta \xi =0$, then one can write equations (\ref{2COMPLEX},\ref{Hyperini variationsCOMPLEX}/\ref{2SPLIT},\ref{Hyperini variationsSPLIT}) as follows
\be
    \left[ {\begin{array}{*{20}c}
   {\frac{1}{2}\left[ {\left( {\partial _M \sigma } \right) - ce^{  \sigma } \left( {\partial _M a + \frac{c}{2}f_M } \right)} \right]\Gamma ^M} & {} & {j\frac{{e^{\frac{\sigma }{2}} }}{{\sqrt 2 }}\left( {\partial _M \chi } \right)\Gamma ^M}  \\
   {} & {} & {}  \\
   {-j\frac{{e^{\frac{\sigma }{2}} }}{{\sqrt 2 }}\left( {\partial _N \bar \chi } \right)\Gamma ^N} & {} & {\frac{1}{2}\left[ {\left( {\partial _N \sigma } \right) + ce^{  \sigma } \left( {\partial _N a - \frac{c}{2}\bar f_N } \right)} \right]\Gamma ^N}  \\
\end{array}} \right]\left( {\begin{array}{*{20}c}
   {\epsilon_1}  \\
   {}  \\
   {\epsilon_2}  \\
\end{array}} \right) = 0,
\ee
satisfied if the determinant of the given matrix vanishes:
\be
    d\sigma  \wedge \star d\sigma  - c^2 e^{2\sigma } \left( {da + \frac{c}{2}f} \right) \wedge \star\left( {da - \frac{c}{2}\bar f} \right) + 2e^\sigma  d\chi  \wedge \star d\bar \chi  = 0.\label{BPS condition}
\ee

Note that (\ref{BPS condition}) is true whether one inserts a ``$j$'' in the $d\chi$ term or not. Using (\ref{BASIC ANSATZ}) and (\ref{BPS condition}) into the dilaton field equation (\ref{Dilaton eomCOMPLEX}/\ref{Dilaton eomSPLIT}) gives
\be
    \left( {\Delta \sigma } \right)\star \mathbf{1} + \frac{1}{2}d\sigma  \wedge \star d\sigma  = \alpha ^2 \left( {1 + \frac{{c^2 }}{2}} \right)e^{ - 2\sigma } dH \wedge \star dH,\label{Dilaton a BPS}
\ee
and the components of the stress tensor also reduce to
\bea
 T_{ab}  &=& \frac{\alpha ^2}{4} \left( {1 + c^2 } \right)\eta _{ab} e^{ - 2\sigma } \left( {\partial _\alpha  H} \right)\left( {\partial ^\alpha  H} \right) \\
 T_{\mu \nu }  &=& \frac{\alpha ^2}{4} \left( {1 + c^2 } \right)g_{\mu \nu } e^{ - 2\sigma } \left( {\partial _\alpha  H} \right)\left( {\partial ^\alpha  H} \right) - \frac{\alpha ^2}{2} \left( {1 + c^2 } \right)e^{ - 2\sigma } \left( {\partial _\mu  H} \right)\left( {\partial _\nu  H} \right).
\eea

Finally, the Einstein equations lead to
\bea
 \frac{1}{2}B\left( {3 - p} \right)\left[ {B\left( {2 - p} \right) - 1} \right]d\sigma  \wedge \star d\sigma  &=&\nonumber\\ \alpha ^2 \left[ {\frac{1}{4}\left( {1 + c^2 } \right)} \right. &-& \left. {\frac{1}{2}B\left( {3 - p} \right)\left( {2 + c^2 } \right)} \right]e^{ - 2\sigma } dH \wedge \star dH \label{BAEqn1}\\
 B\left( {2 - p} \right)\left( {2B + 1} \right)d\sigma  \wedge \star d\sigma  &=& \alpha ^2 \left[ {B\left( {2 - p} \right)\left( {2 + c^2 } \right) - \left( {1 + c^2 } \right)} \right]e^{ - 2\sigma } dH \wedge \star dH \label{BAEqn2}\\
 B\left( {2 - p} \right)\left[ {B\left( {1 - p} \right) - 1} \right]d\sigma  \wedge \star d\sigma  &=&\nonumber\\ \alpha ^2 \left[ {\frac{1}{2}\left( {1 + c^2 } \right)} \right. &-& \left. {B\left( {2 - p} \right)\left( {2 + c^2 } \right)} \right]e^{ - 2\sigma } dH \wedge \star dH,\label{BAEqn3}
\eea
where (\ref{BASIC ANSATZ}, \ref{BPS condition}) were used.

To begin, we check that for $c=i$, equations (\ref{Dilaton a BPS}, \ref{BAEqn1}, \ref{BAEqn2}, \ref{BAEqn3}) are satisfied only for the instanton case $B=0$ ($\alpha=0$ and $\alpha\ne 0$) as well as for the ($p=3$, $\alpha=0$) case where one can show that $B=-1/2$. Attempts to find a solution for the most general $p$-brane case with non-vanishing $\alpha$ fail for most $p$ cases and are problematic for $p=2$ and $p=3$. For the former, one finds that the conditions $\left( {\Delta \sigma } \right)\star \mathbf{1} = 0$ and $d\sigma  \wedge \star d\sigma  = \alpha ^2 e^{ - 2\sigma } dH \wedge \star dH$ have to be simultaneously true, which is not possible within the symmetries assumed. For the latter, we also find that $\left( {\Delta \sigma } \right)\star \mathbf{1} = Bd\sigma  \wedge \star d\sigma $ and $\left( {2B + 1} \right)d\sigma  \wedge \star d\sigma  = \alpha ^2 e^{ - 2\sigma } dH \wedge \star dH$ have to be simultaneously satisfied. This can be easily shown to be possible if and only if $\alpha$ is imaginary, which is only true for the instanton case. We will show below that even the allowed cases pose serious problems with the BPS condition $\delta \xi = 0$, and one is forced to abandon the complex representation altogether.

For the case $c=j$, there are only two allowed solutions: the instanton ($B=0$, $\alpha=0$) and the 3-brane ($B=-1/2$, $\alpha=0$). All non-vanishing $\alpha$ cases do not simultaneously satisfy (\ref{Dilaton a BPS}, \ref{BAEqn1}, \ref{BAEqn2}, \ref{BAEqn3}). As we will show below the allowed cases smoothly satisfy the vanishing of the SUSY variations condition and enforce the argument that the split-complex representation of the UV fields is the more natural representation to use.

\section{Non-BPS \emph{almost}-solutions in the complex representation}

In this section, we give all possible solutions found for the ordinary complex form of the theory and show that, in addition to some reality issues, these solutions cannot possible satisfy the BPS condition $\delta \xi=0$. This can be traced back to the ambiguous sign problem of (\ref{2COMPLEX}) and (\ref{Hyperini variationsCOMPLEX}).

\subsection{Instantons with vanishing $\alpha$:}

For all $\alpha=0$ cases equation (\ref{Dilaton a BPS}) becomes
\be
    \left( {\Delta \sigma } \right)\star \mathbf{1}  + \frac{1}{2}d\sigma  \wedge \star d\sigma  = 0\,\,\,\,\,\,\,\,\,\, \to \,\,\,\,\,\,\,\,\,\,\Delta e^{{\sigma  \mathord{\left/ {\vphantom {\sigma  2}} \right. \kern-\nulldelimiterspace} 2}}  = 0.
\ee

Equivalently
\be
    \sigma  = 2\ln \left( { H} \right).\label{InstantonALPHA=01}
\ee

The $\left(\chi, \bar\chi\right)$ equations (\ref{Chi eomCOMPLEX}, \ref{Chi bar eomCOMPLEX}) are easily integrated to give
\bea
 \chi  &=&  n - qe^{{{ - \sigma } \mathord{\left/
 {\vphantom {{ - \sigma } 2}} \right.
 \kern-\nulldelimiterspace} 2}}   \nonumber\\
 \bar \chi  &=&  \bar n - \bar qe^{{{ - \sigma } \mathord{\left/
 {\vphantom {{ - \sigma } 2}} \right.
 \kern-\nulldelimiterspace} 2}}  ,\label{InstantonALPHA=02}
\eea
where $q$ and $n$ are arbitrary complex constants. From this one can easily find $da$ using (\ref{BASIC ANSATZ}) with $\alpha=0$ and integrate to find
\be
    a =  k - {\mathop{\rm Im}\nolimits} \left( {n\bar q} \right)e^{{{ - \sigma } \mathord{\left/
 {\vphantom {{ - \sigma } 2}} \right.
 \kern-\nulldelimiterspace} 2}}  ,\label{InstantonALPHA=03}
\ee
where $k$ is a real integration constant. However, a problem arises when one tries to verify the BPS condition (\ref{BPS condition}), since it yields
\be
    \left| q \right|^2  =  - 2.\label{InstantonALPHA=04}
\ee

Clearly, this condition cannot be satisfied unless $q$ is split-complex. So one concludes that only non-BPS instantons coupled to the above fields can exist for as long as we insist on using a pure complex representation of the theory.

\subsection{Instantons with non-vanishing $\alpha$:}

Equation (\ref{Dilaton a BPS}) gives
\be
    \left( {\Delta \sigma } \right)\star \mathbf{1} + \frac{1}{2}d\sigma  \wedge \star d\sigma  = \,\frac{{\alpha ^2 }}{2}\,e^{ - 2\sigma } dH \wedge \star dH,
\ee
which is solved by
\be
    \sigma  = \ln \left( { H} \right),\label{specialsigma}
\ee
only if $\alpha=i$ as expected. The $\left(\chi, \bar\chi\right)$ equations (\ref{Chi eomCOMPLEX}, \ref{Chi bar eomCOMPLEX}) can be integrated once to give:
\bea
 e^\sigma  d\chi  -  \chi dH = qdH \nonumber\\
 e^\sigma  d\bar \chi  +  \bar \chi dH = \bar qdH,\label{InsertjinCHI}
\eea
clearly \emph{not} complex conjugates, once again due to a sign difference. Integrating these two first order differential equations gives the strange solution
\bea
 \chi  &=& \varphi  e^\sigma   - q \nonumber\\
 \bar \chi  &=& \bar \varphi e^{ - \sigma }  + \bar q,
\eea
where $\varphi $ is a complex integration constant. This solution is reminiscent of, albeit simpler than, the one found in \cite{Gutperle:2000sb}, in that it has the strange property $\chi  \propto {1 \mathord{\left/  {\vphantom {1 {\bar \chi }}} \right.  \kern-\nulldelimiterspace} {\bar \chi }}$ such that $\chi$ and $\bar \chi$ are not complex conjugates. As shown in the same source, making the $\sigma$ solution (\ref{specialsigma}) more general partially improves the situation by making $\chi$ and $\bar \chi$ complex conjugates at radial infinity. The more serious problem, however, arises when once again one tries to check the BPS condition (\ref{BPS condition}) to find that it is not satisfied unless
\be
    1 + \alpha ^2  = 2\left| {\varphi } \right|^2 e^\sigma,
\ee
which, along with $\alpha=i$, can only be true if and only if $\left| {\varphi} \right|^2 $ is zero, only possible if $\varphi$ is a null split-complex number. As such, it would be reminiscent of the result we found in \cite{Emam:2007qa}.

\subsection{3-branes with vanishing $\alpha$:}

This is the only $p$-brane case where equations (\ref{BAEqn1}, \ref{BAEqn2}, \ref{BAEqn3}) are satisfied. They give $B =  - {1 \mathord{\left/ {\vphantom {1 2}} \right. \kern-\nulldelimiterspace} 2}$. Because of the vanishing of $\alpha$, the solutions (\ref{InstantonALPHA=01}), (\ref{InstantonALPHA=02}) and (\ref{InstantonALPHA=03}) are satisfied here as well. Unfortunately, so is the result (\ref{InstantonALPHA=04}) which once again implies a non-BPS solution.

\section{Exact BPS solutions in the split-complex representation}

Since only solutions with $\alpha=0$ are found in this case, the UH fields have a similar form to that of (\ref{InstantonALPHA=01}), (\ref{InstantonALPHA=02}) and (\ref{InstantonALPHA=03}). They are well-behaved and do not suffer from the issues discussed in the past section (specifically, equation (\ref{InstantonALPHA=04}) is no longer a problem). The two split-complex solutions differ only in their metrics as well as the explicit spatial dependence of the harmonic function $H$.

\subsection{Instantons with vanishing $\alpha$:}

The harmonic function for the instanton case is simply
\be
    H = \frac{{Q_I }}{{r^3 }} + h,\label{H_I}
\ee
where $Q_I$ is an arbitrary charge and the constant $h$ is defined such that the value of the dilaton at radial infinity is $\sigma _\infty   = 2\ln \left( h \right)$. Using (\ref{InstantonALPHA=01}), (\ref{InstantonALPHA=02}) and (\ref{InstantonALPHA=03}), the complete solution in the $D=5$ Euclidean background is then
\bea
 \sigma  &=& 2\ln \left( { \frac{{Q_I }}{{r^3 }} +  h} \right) \\
 \chi  &=& \chi _0  - \frac{{q r^3 }}{{\left( {Q_I  + hr^3 } \right)}} \\
 \bar \chi  &=& \bar \chi _0  - \frac{{\bar q r^3 }}{{\left( {Q_I  + hr^3 } \right)}} \\
 a &=& a_0  - \frac{{{{\mathop{\rm Im}\nolimits} \left( {\bar q \chi _\infty  } \right)} r^3 }}{{\left( {Q_I  + hr^3 } \right)}},
\eea
where $\chi _0 $ and $a_0 $ are the $\left(r=0\right)$ values of these fields, and ${\chi _\infty  }$ is the $\left(r\rightarrow \infty\right)$ value of $\chi$. The following relations between the various constants hold:
\be
 h\left( {\chi _0  - \chi _\infty  } \right) = q \quad\quad {\rm and}\quad\quad
 \left| q \right|^2  =  - 2.
\ee

Finally, the vanishing of the SUSY variations (\ref{1SPLIT}) and (\ref{Gravitini variationsSPLIT}) gives
\be
   \epsilon _s  = e^{ - \frac{\sigma }{{4\sqrt 2 }}} \hat \epsilon _s ,\,\,\,\,\,\,\,\,s = 1,2,\label{Spinors}
\ee
where $\hat \epsilon _s$ are constant spinors.

\subsection{3-branes with vanishing $\alpha$:}

For the 3-brane ($B =  - {1 \mathord{\left/ {\vphantom {1 2}} \right. \kern-\nulldelimiterspace} 2}$) the function $H$ is harmonic in the single transverse direction $r=x^4$ and has the form
\be
    H\left( r \right) = Q_3 \sqrt r.\label{H_3}
\ee

This gives the metric
\bea
 ds^2  &=& \eta _{ab} dx^a dx^b  + e^{ - \sigma } dr^2 ,\,\,\,\,\,\,\,\,\,\,\,\,\,a,b = 0,1,2,3 \nonumber\\
  &=& \eta _{ab} dx^a dx^b  + \frac{{dr^2 }}{{ Q^2_3 r}},
\eea
and the UH fields
\bea
 \sigma  &=& 2\ln \left( { Q_3 \sqrt r } \right) \nonumber\\
 \chi  &=& \chi _\infty   - \frac{q}{{ Q_3 \sqrt r }} \nonumber\\
 \bar \chi  &=& \bar \chi _\infty   - \frac{{\bar q}}{{ Q_3 \sqrt r }} \nonumber\\
 a &=& a_\infty   - \frac{{{\mathop{\rm Im}\nolimits} \left( {\chi _\infty  \bar q} \right)}}{{ Q_3 \sqrt r }}.
\eea

One can easily check that the SUSY spinors $\epsilon_s$ have exactly the same form as (\ref{Spinors}). Note the interesting property, possibly worth exploring in future research, that in the far field region $\left(r\rightarrow \infty\right)$, the metric reduces to four dimensional Minkowski. In other words, the fifth dimension is lost. Attempts to fix this by adding a constant to (\ref{H_3}), as in the previous case (\ref{H_I}), violated the UH field equations.


\section{Conclusion}

We argued that, in general, the most natural way to represent the fields of the universal hypermultiplet, specifically the axions, is with the split-complex representation. In addition to past evidence, we enforced this argument by attempting to classify all possible solutions based on the general ansatz (\ref{BASIC ANSATZ}) and found that BPS solutions cannot exist as long as one insists on using the ordinary complex representation. On the other hand, switching to the split-complex form of the theory, all problems are instantly fixed and we find two well-behaved solutions representing BPS instantons and 3-branes. This argument may be thought of as a special case of \cite{Emam:2009xj} where we had argued that the same theory, but with the full set of non-trivial hypermultiplet scalars, is best represented using a formulation based on the theory's inherent symplectic structure. This was done by defining a symplectic space on which the hypermultiplet fields appear as either vectors or scalars. The connection lies in the fact that the inner product of vectors on this space has exactly the same form as the modulus of split-complex numbers (\ref{Hyperbolic scalar product}). The classification of hypermultiplet solutions is an ongoing quest that may further be made rigorous by adopting the split-complex representation. Possible directions of future research include generalizing the background metric to include other possible configurations of branes. It may also be worthwhile to further explore the connection with the more general symplectic structure of the full theory.

\pagebreak

\end{document}